\documentclass[a4paper,11pt]{article}
\pdfoutput=1 

\usepackage{jheppub} 

\usepackage[T1]{fontenc} 
\usepackage{amsmath}
\usepackage{slashed}
\usepackage{enumitem}

\title{\boldmath Electromagnetic Albedo of Quantum Black Holes}

\author[a,b,d,1]{Wan Zhen Chua,\note{Corresponding author.}}
\author[a,b,c]{Niayesh Afshordi}

\affiliation[a]{Perimeter Institute For Theoretical Physics, 31 Caroline St N, Waterloo, ON N2L 2Y5, Canada}
\affiliation[b]{Department of Physics and Astronomy, University of Waterloo, Waterloo, ON N2L 3G1, Canada}
\affiliation[c]{Waterloo Centre for Astrophysics, University of Waterloo, Waterloo, ON N2L 3G1, Canada}
\affiliation[d]{Department of Physics, Cornell University, Ithaca, New York, USA}

\emailAdd{wc646@cornell.edu}
\emailAdd{nafshordi@pitp.ca}

\abstract{We compute the albedo (or reflectivity) of electromagnetic waves off the electron-positron Hawking plasma that surrounds the horizon of a Quantum Black Hole. We adopt the ``modified firewall conjecture'' for fuzzballs  \cite{Mathur:2005zp,Guo:2017jmi}, where we consider significant electromagnetic interaction around the horizon.  While prior work has treated this problem as an electron-photon scattering process, we find that the incoming quanta interact {\it collectively} with the fermionic excitations of the Hawking plasma at low energies. We derive this via two different methods: one using relativistic plasma dispersion relation, and another using the one-loop correction to photon propagator. Both methods find that the reflectivity of long wavelength photons off the Hawking plasma is significant, contrary to previous claims.  This leads to the enhancement of the electromagnetic albedo for frequencies comparable to the Hawking temperature of black hole horizons in vacuum. We comment on possible observable consequences of this effect.}

\begin{document} 
\maketitle
\flushbottom

\section{Introduction}
\label{sec:intro}
From the theory of General Relativity, black hole is a spacetime with vacuum region around its horizon and mass concentrated at the singularity. Classically, no signal can be transmitted to an outside observer since nothing can propagate faster than the speed of light. However, it is well known that General Relativity breaks down near the energy of Planck scale, and thus quantum effects may modify the near horizon behavior drastically. 

In order to describe quantum black holes, a number of conjectures have been proposed (e.g., \cite{PrescodWeinstein:2009mp,Saravani:2012is,Giddings:2017mym,Kawai:2017txu,Oshita:2018fqu,Wang:2018gin,Abedi:2020ujo,Abedi:2018npz,Abedi:2020sgg}), including black holes being horizonless (fuzzballs) \cite{Mathur:2005zp}. Horizonless microstate constituents solve the information paradox \cite{Hawking:1974sw,Hawking:1976ra,Almheiri:2012rt,Mathur} by allowing black holes to emit blackbody radiation and no information is trapped within. Despite the vast difference in the microscopic details of the proposed conjectures, the Bekenstein-Hawking entropy \cite{Bekenstein:1972tm,Bekenstein:1973ur,Gibbons:1976ue} is a universal macroscopic thermodynamic quantity agreed by all. However, another potentially universal property, i.e. the surface reflectivity, was first proposed in \cite{Kuchiev:2003rv,Kuchiev:2005em,Oshita:2019sat,Wang:2019rcf}. Dissipative effects near the horizon were considered in \cite{Oshita:2019sat,Wang:2019rcf}, who proposed that the surface reflectivity exhibits Boltzmann suppression at high frequencies. However, from the perspective of quantum gravity, there is no unique way of adding a viscous term to the dispersion relation that accounts for dissipation. 

As an alternative to the consideration of dissipative effects near the horizon, we consider the dispersion relation of a relativistic electron-positron plasma \cite{Medvedev:1999pt,Thoma:2008tb} in Rindler coordinates (Hawking/Unruh plasma). The idea of having a Hawking plasma as a near horizon behavior was introduced in \cite{Mathur:2005zp,Guo:2017jmi} through the ``modified firewall conjecture''. It is proposed in \cite{Guo:2017jmi} that long wavelength modes can indeed be scattered off from black holes, however, of negligible probability. We dispute this claim by considering a collective fermionic interaction instead of treating it as a standard scattering process.

In Section \ref{sec:fuzzball}, we review the fuzzball proposal that is discussed extensively in \cite{Mathur:2005zp,Guo:2017jmi}. In Section \ref{sec:dispersion}, using the standard dispersion relation of photons in a relativistic plasma, we obtain the flux reflectivity from the ratio of the amplitude of outgoing and incoming waves in Rindler space. In Section \ref{sec:1-loop}, we directly obtain the flux reflectivity by projecting the QED one-loop correction of the photon propagator into Rindler modes. The results in Sections \ref{sec:dispersion} and \ref{sec:1-loop} show the same frequency dependence of reflectivity/albedo, and only differ by an ${\cal O}(1)$ factor, a potential artefact of approximations in each approach. 
We then conclude our work in Section \ref{sec:conclusions}. 

Throughout the paper, we have adopted the units $\hbar = c = k_B = 1$. 
\section{Review on Fuzzballs}
\label{sec:fuzzball}
The fuzzball proposal is discussed in great detail in \cite{Guo:2017jmi,Mathur:2005zp}. In this section, we specifically review the ``modified firewall behavior'' that we adopt in our calculation. 

The fuzzball proposal states that black holes are composed of microstates that do not possess horizons. This implies fuzzballs radiate information like a blackbody in the absence of Hawking radiation, leading to a resolution of information paradox. It is proposed that the quantum behavior of fuzzballs is hard to observe, which we shall later dispute in this paper. 

\begin{figure}[tbph]
	\centering
	\includegraphics[width=0.55\textwidth]{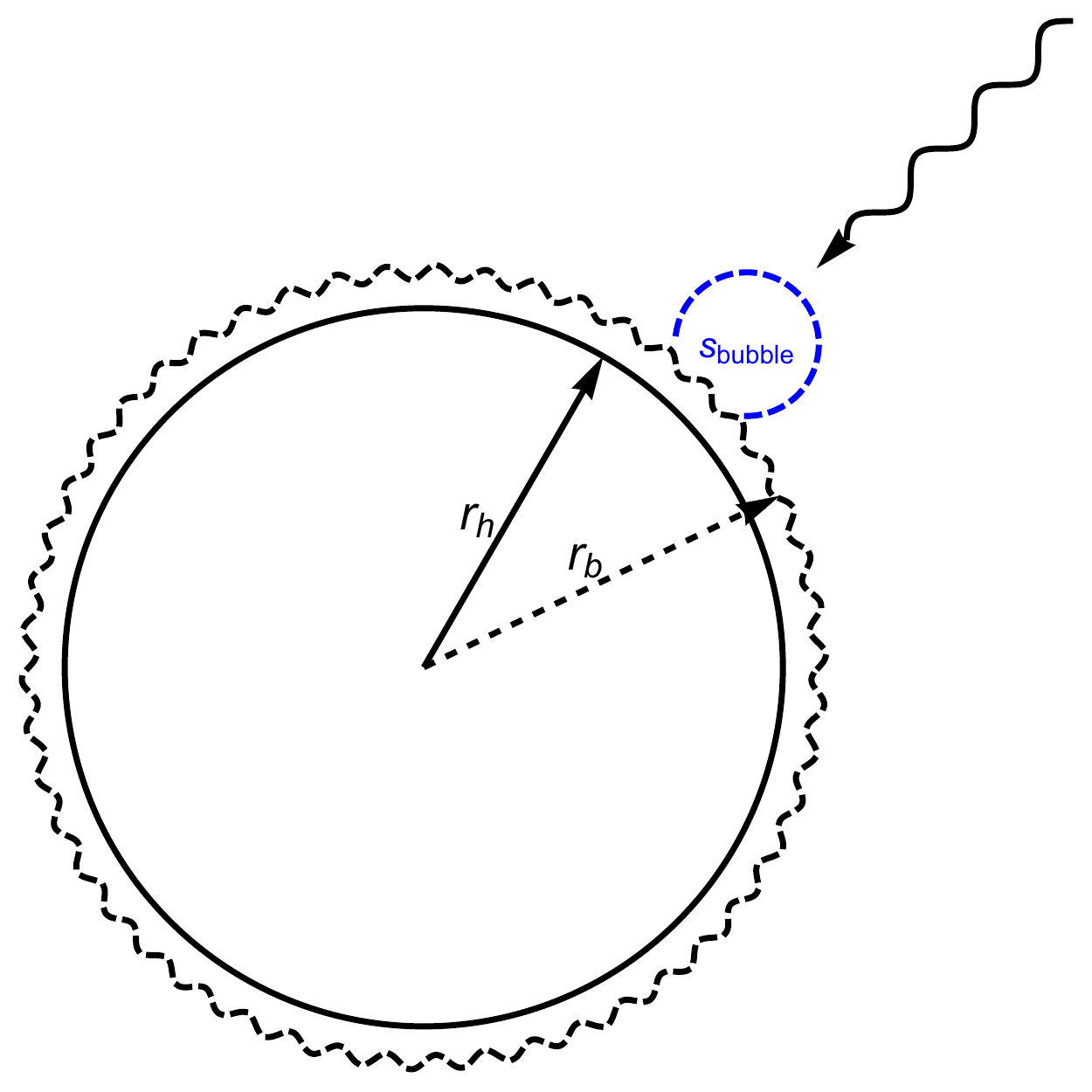}
	\caption{Figure shows the deformation of the fuzzball surface due to backreaction. The deformation is local and a ``bubble'' is formed. $r_h$ is the classical horizon radius and $r_b$ is the boundary of the fuzzball surface. }
	\label{Fig:sbubble}
\end{figure}

In \cite{Guo:2017jmi}, the ``modified firewall behavior'' is proposed to allow black holes radiate information without violating causality. Due to the backreaction caused by energy of the infalling object, we have a ``bubble'' formed locally with radius $s_{\text{bubble}}$ from the fuzzball surface, as shown in Figure \ref{Fig:sbubble}. In the absence of significant interaction, the infalling object would be ``engulfed'' by this new horizon.

Quantum gravitational effects should arise at some proper distance $s>s_{\text{bubble}}$ to allow electromagnetic reflection. Applying semiclassical physics to the region $s>s_{\text{bubble}}$, we investigate interactions of the infalling object with the emitted radiation from the fuzzball surface. We come to a conclusion that if $P_{\text{interact}} \sim 1$, there is ``modified firewall behavior''. On the other hand, if $P_{\text{interact}} \ll 1$, the ``modified firewall behavior'' is absent.

Let us provide a rough estimate to $s_{\text{bubble}}$. In reality, the ``bubble'' is not spherically symmetric. However, we neglect this deformation in our rough calculation. We have the thermodynamical relation for the change in entropy when the fuzzball is deformed
\begin{equation}
    \delta S_{\text{Bek}} = \frac{E}{T}~,
\end{equation}
where $E$ is the backreacted energy and applying the Bekenstein-Hawking entropy relation $\delta S_{\text{bek}} = \delta A/G$ where $\delta A \sim s_{\text{bubble}}^2$
\begin{equation}
\begin{split}
    \frac{s_{\text{bubble}}^2}{l_p^2} &\sim \frac{E}{T}~, \\
    s_{\text{bubble}} &\sim \left(\frac{E}{T}\right)^{1/2}l_p~.
    \end{split}
\end{equation}

Next, we provide a rough estimation on the scattering energy domain of the incoming quanta. The local temperature and photon number density in the orthonormal frame are given by \cite{Guo:2017jmi}
\begin{equation}\label{eq:Tn}
    \hat{T} \sim \frac{1}{s} ~, \quad \hat{n} \sim \frac{1}{s^3}~,
\end{equation}
respectively. By taking $\theta=\pi/2$, the orthonormal frame can be expressed as
\begin{equation}
    d\hat{t} = \left(1 - \frac{r_h}{r}\right)^{1/2}dt~, ~~d\hat{r} = \left(\frac{1}{1-\frac{r_h}{r}}\right)^{1/2}dr~, ~~d\hat{\theta} = r d\theta~,~~ d\hat{\phi} = r d\phi~.
\end{equation}
The energy of a radiated photon is given by
\begin{equation}
    \hat{E}_{\text{radiation}} \sim \hat{T} \sim \frac{1}{s}~,
\end{equation}
and the energy of the infalling particle in the local orthonormal frame is given by
\begin{equation}
    \hat{E}_{\text{infalling}} \sim (-g_{tt})^{-1/2}E \sim \frac{r_h}{s}E \sim \left(\frac{E}{T_H}\right)\frac{1}{s}~,
\end{equation}
where we have taken the temperature $T_H$ to scale as $\sim 1/r_h$ and the proper distance $s$ to be
\begin{equation}\label{eq:proper}
    s = \int_{r_h}^r \frac{dr'}{\left(1-\frac{r_h}{r'}\right)^{1/2}} \approx 2 r_h^{1/2} (r-r_h)^{1/2}~.
\end{equation}
In the centre of mass frame \begin{equation}\label{eq:Ecm}
    \hat{E}_{\text{cm}} \sim \sqrt{\hat{E}_{\text{radiation}}\hat{E}_{\text{infalling}}} \sim \frac{1}{s}\sqrt{\frac{E}{T_H}}~.
\end{equation} 

For electron-photon scattering, the cross section is given by
\begin{equation}\label{eq:sigma}
    \sigma_{e\gamma} \sim \frac{\alpha^2}{\hat{E}_{\text{cm}}^2} \sim \frac{\alpha^2s^2T_H}{E}~,
\end{equation}
where $\alpha$ is the fine structure constant. Using \eqref{eq:Tn} and \eqref{eq:sigma}, the differential probability of interaction is then given by
\begin{equation}
    \frac{dP_{\text{interact}}^{e\gamma}}{ds} \sim \sigma_{e\gamma}\hat{n} \sim \frac{\alpha^2}{\hat{E}_{\text{cm}}^2s^3} \sim \frac{\alpha^2T_H}{Es}~.
\end{equation}
Setting the inital proper distance to be of order $\sim r_h$ and integrating it gives
\begin{equation}
    P_{\text{interact}}^{e\gamma}(s) \sim \alpha^2 \frac{T_H}{E}\ln \frac{1}{s T_H}~.
\end{equation}
Requiring $P_{\text{interact}}(s) \sim 1$ gives the proper distance to the horizon of black hole ``photosphere'':
\begin{equation}
    s_{\text{interact}}^{e\gamma} \sim \frac{1}{T_H} \exp\left(-\frac{ E}{\alpha^2 T_H}\right)~.
\end{equation}
For the existence of modified firewall behavior, we require $s_{\text{interact}}^{e\gamma} \gg s_{\text{bubble}}$
\begin{equation}\label{eq:rhE}
\begin{split}
    \frac{1}{T_H} \exp\left(-\frac{ E}{\alpha^2 T_H}\right) &\gg \sqrt{\frac{E}{T_H}}l_p~, \\
    \frac{ E}{T_H} &\ll \alpha^2 \ln \left(\sqrt{\frac{1}{E T_H}} \frac{1}{l_p}\right)~.
    \end{split}
\end{equation}
As a rough estimation, even setting the right hand side of \eqref{eq:rhE} to be of order one, we have
\begin{equation}\label{eq:rh}
    E \ll T_H~,
\end{equation}
which shows that the wavelength must dominate over the horizon radius for the existence of ``modified firewall behavior''.

We now consider the situation where an electron is at rest and scattered off by an incoming photon. The scattering cross section is given by
\begin{equation}
    \sigma \sim \frac{\alpha^2}{m_e^2}~, ~~\omega \leq m_e~, 
\end{equation}
and the probability for interaction is given by
\begin{equation}
    P_{\text{interact}} \sim s \frac{dP_{\text{interact}}}{ds} \sim s\sigma\hat{n} \sim \frac{\alpha^2}{m_e^2 s^2}~.
\end{equation}
In the orthonormal frame, we have $m_e \sim 1/s$ and hence, we conclude
\begin{equation}
    P_{\text{interact}} \sim \alpha^2 \sim 10^{-4}~. 
\end{equation}
In addition to the interaction probability $P_{\text{interact}} \sim 10^{-4}$, we have to consider the probability of the photon escaping to infinity since the scattering angle of the photon is assumed to randomised. We have the following approximation for the angle of escape
\begin{equation}
    \sin \Phi \sim \Phi \sim \frac{s}{r_h} \sim \frac{T_H}{m_e}~,
\end{equation}
and this leads to the probability of emergence
\begin{equation}
    P_{\text{emergence}} \sim \Phi^2 \sim \frac{T_H^2}{m_e^2}~.
\end{equation}
Combining the two probabilities, the overall probability for scattering the photon off the Hawking plasma is given by
\begin{equation}
    {\rm Reflection~Probability} \sim P_{\text{interact}}P_{\text{emergence}} \sim \frac{\alpha^2 T_H^2}{m_e^2} \ll 1~,
\end{equation}
showing the absence of ``modified firewall behavior''. In Section \ref{sec:dispersion} and \ref{sec:1-loop}, we re-examine this claim by considering collective interaction of fermions with the incoming photon by approximately solving the scalar wave equation near the horizon. Notably, we find that $m_e$ ultimately drops out of the calculations. 

\section{Dispersion Relation of EM Waves in the Hawking Plasma}
\label{sec:dispersion}
The relativistic plasma involves collective plasma modes of fermions and are created when the plasma is of very high temperature, i.e., thermal energy $T$ of plasma excitations is much larger than rest mass of plasma particles. The dispersion relation for relativistic plasma takes the following exact form \cite{Medvedev:1999pt,Thoma:2008tb}
\begin{equation}\label{eq:dispersion_full}
    \frac{k^2}{\omega^2} = 1 + \frac{3}{4}\frac{\omega_p^2}{\omega k}\left[\left(1-\frac{\omega^2}{k^2}\right)\ln \bigg{|}\frac{\omega-k}{\omega+k}\bigg{|}-\frac{2 \omega}{k}\right]~,
\end{equation}
where $\omega_p$ is the plasma frequency. When the incoming photon is coupled strongly via the oscillating magnetic and electric fields with the plasma particles, we interpret the blueshift in frequency as the incoming photon gaining an effective mass $m_\gamma$. We apply the following approximation to the dispersion relation \cite{Medvedev:1999pt}
\begin{equation}\label{eq:dispersion}
    \omega^2 \simeq k^2 + m_{\gamma}^2~.
\end{equation}
\eqref{eq:dispersion} is valid as long as $\omega \simeq k \gg m_\gamma$ and this relation is exact in the non-relativistic case. 
The two limiting regimes of the effective photon mass in the relativistic plasma is given by:
\begin{eqnarray}\label{eq:massphoton}
     && m_{\gamma}^2 \simeq \frac{3}{2}\omega_p^2 \simeq \frac{e^2 T^2}{6}~, ~~(T \gg m_e)~,\\
    && m_{\gamma}^2 \simeq \frac{ e^2 (n_{e^+}+n_{e^-})}{m_e} = \frac{ \sqrt{2} e^2}{m_e}\left(\frac{m_e T}{\pi}\right)^{3/2}e^{-m_e/T}~,  ~~(T \ll m_e)~.
\end{eqnarray}
In the non-relativistic regime, the effective photon mass is obtained using classical transport theory:
\begin{equation}
    n_{e^{\pm}} = \frac{g}{(2 \pi)^3}\int d^3 p \frac{1}{e^{\sqrt{p^2 + m_e^2}/T}+1} \simeq \frac{1}{\sqrt{2}}\left(\frac{m_e T}{\pi}\right)^{3/2}e^{-m_e/T}~,
\end{equation}
where $g=2$. 

\begin{figure}[tbph]
	\centering
	\includegraphics[width=0.8\textwidth]{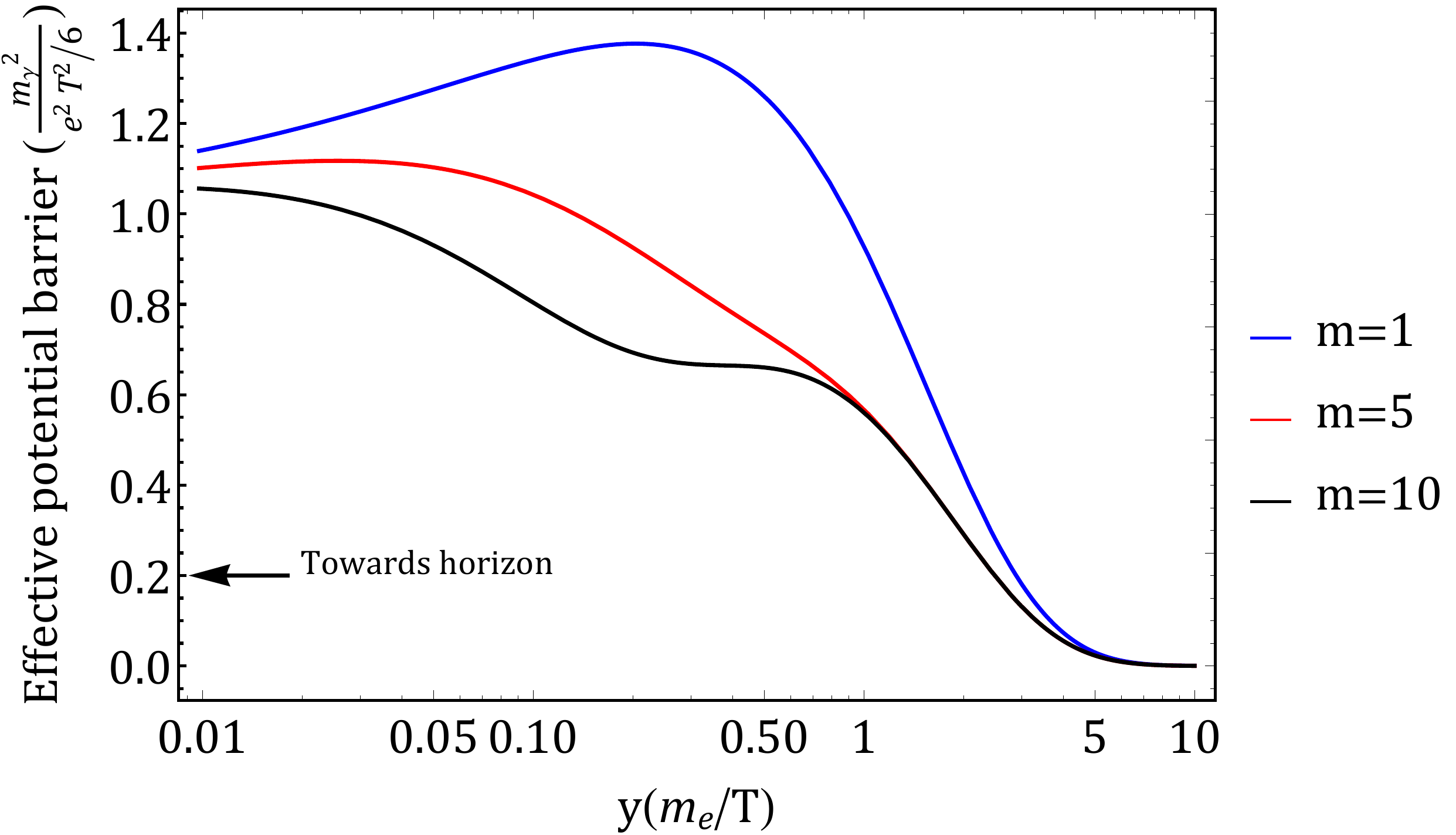}
	\caption{Some possible interpolations of the effective potential barrier for photons in \eqref{eq:effectmass}. Here, $y=m_e/T$, is defined in terms of blueshifted Hawking temperature. The black hole horizon lies at $y \to 0$, or $T \to \infty$. }
	\label{Fig:effectmass}
\end{figure}

We are interested in mode functions near the horizon as that is where reflection occurs. However, the interactions between the EM waves and the Hawking plasma allows us to neglect the gravitational effects of the black hole. The exterior metric in the $(T,X)$ plane can be approximated as Rindler
\begin{equation}\label{eq:rindler}
    ds^2 = e^{2 \kappa x}\left(-dt^2 + dx^2\right)+dY^2+dZ^2~,
\end{equation}
where $\kappa = 2 \pi T_H$ is the surface gravity and it describes the Hawking plasma in the accelerated frame. In principle, we have to solve for the collisionless Boltzmann equation and Maxwell equations in Rindler to obtain the dispersion relation (as shown in Appendix \ref{appendix:CBE_Rindler}). However, we weren't able to obtain plane wave solutions at the first order approximation and we chose to introduce an interpolation that describes the incoming photon experiencing a transition from the non-relativistic to the relativistic regime of the effective photon mass in Rindler space. A simple possible interpolating function can be written
\begin{equation}\label{eq:effectmass}
    m^2_\gamma \simeq \frac{e^2T^2}{6}\left(C_R+ \frac{6 \sqrt{2m_e}}{\sqrt{\pi^3 T}}e^{-m_e/T}\right)~,
\end{equation}
where  $C_R$ is a function of temperature that ensures smooth interpolation between the ultra-relativistic and non-relativistic regimes. We see that upon approaching the Hawking plasma, $T$ dominates over $m_e$ and the frequency of the photon is blueshifted significantly. We can choose the coordinates such that the photon is travelling in the $t-x$ plane upon approaching the plasma. $T$ in \eqref{eq:effectmass} is the effective temperature given by
\begin{equation}\label{eq:efftemp}
    T = T_H \exp(-\kappa x) = \frac{\kappa}{2 \pi}\exp(-\kappa x)~.
\end{equation}
With the massive Klein-Gordon equation\footnote{Here, $\psi$ could be taken as  $E_Y$ or $E_Z$, i.e. the transverse components of the electric field, while propagation is radial, i.e. along X direction, and $g_{\mu\nu}$ only refers to the 2D Rindler geometry in the T-X plane.} and effective photon mass relation \eqref{eq:effectmass}, we have the following wave equation
\begin{eqnarray}\label{eq:wave}
&& \frac{1}{\sqrt{-g}}\partial_\mu (\sqrt{-g} g^{\mu\nu} \partial_\nu \psi(x)) - m_\gamma^2 \psi(x) = 0~, \\
   && \frac{d^2 \psi(x)}{dx^2} + \omega^2 \psi(x) - \frac{e^2 T_H^2}{6}\left(C_{\rm R}+\frac{6 \sqrt{2m_e}}{\sqrt{\pi^3 T(x)}} e^{-m_e/T(x)} \right)\psi(x) = 0~.
\end{eqnarray}
Making a change of variable $y=m_e/T(x)= m_e\exp(\kappa x)/T_H$,  yields
\begin{equation}\label{eq:wave_y}
    y^2\frac{d^2\psi}{dy^2}+y\frac{d\psi}{dy}+\left[\mu^2-\frac{\alpha}{6\pi}\left(C_R(y)+6 \sqrt{2y\over \pi^3}\exp(-y)\right)\right]\psi=0~,
\end{equation}
where $\alpha\equiv e^2/4 \pi$ is the fine structure constant in natural units and $\mu=\omega/(2\pi T_H) = \omega/\kappa$. A class of interpolating functions allows $C_{\rm R}(y) $ to take the form of $C_{\rm R}(y) = e^{-|m| y}$ where $m \geq 1$ for which $C_{\rm R}$ satisfies the appropriate asymptotic behavior: $C_{\rm R} \rightarrow 1$ as $y \rightarrow 0$ and $C_{\rm R} \rightarrow 0$ such that non-relativistic regime dominates as $y \rightarrow \infty$. In Figure \ref{Fig:effectmass}, we illustrate some possible interpolations of the effective potential barrier for the incoming photon. 
The horizon is at $y \to 0$ and spatial infinity is at $y \to \infty$. The 0th-order ingoing solution in $\alpha$ is given by $A_{\rm in} \exp(-i\omega x) \propto y^{-i\mu}$. The first-order correction in $\alpha$ to the scalar wave equation is given by
\begin{equation}
    y^2\frac{d^2\psi^{(1)}}{dy^2}+y\frac{d\psi^{(1)}}{dy}+\mu^2\psi^{(1)} -\frac{\alpha}{6\pi}\left(e^{-|m| y}+6 \sqrt{2y\over \pi^3}\exp(-y)\right)\psi^{(0)}=0~,
\end{equation}
where the $0$th-order ingoing solution $\psi^{(0)}$ is given by $y^{-i \mu}$. Far away from the horizon ($y \gg 1$)
\begin{equation}
    \psi_{\text{far}} \rightarrow \left(\frac{C_1}{2}-\frac{i C_2}{2}\right)y^{i \mu}+\left(\frac{C_1}{2}+\frac{i C_2}{2}+\frac{i \alpha}{\sqrt{2} \pi^2 \mu}\right)y^{-i \mu}
\end{equation}
We can set
\begin{equation}\label{eq:c1c2_a}
   \frac{C_1}{2}+\frac{i C_2}{2}+\frac{i \alpha}{\sqrt{2} \pi^2 \mu} = 1~,
\end{equation}
such that the flux reflectivity is given by
\begin{equation}
    \mathcal{R}_{\text{QED}} = \frac{1}{4}(C_1^2+C_2^2)~.
\end{equation}
Near the horizon
\begin{equation}
   \psi_{\text{near}} \rightarrow A y^{i \mu}+T_{\text{QED}}y^{-i \mu}~,
\end{equation}
where $A$ is a constant to be set to zero to ensure ingoing boundary condition and $\mathcal{T}_{\text{QED}} = |T_{\text{QED}}|^2$ is the transmission amplitude such that $\mathcal{T}_{\text{QED}} + \mathcal{R}_{\text{QED}} = 1 $. 
Assuming ingoing boundary condition
\begin{equation}\label{eq:c1c2_b}
   A=\frac{C_1}{2}-\frac{i C_2}{2}+\frac{i \alpha}{\sqrt{2} \pi^{5/2} \mu}\Gamma\left(\frac{1}{2}-2 i \mu\right)+\frac{i \alpha|m|^{2 i \mu}}{12 \pi \mu}\Gamma(-2 i \mu) = 0~.
\end{equation}
For a coherent scattering problem, similar to the Hawking radiation, the outgoing flux at infinity is related to the outgoing flux at the horizon by the standard greybody factor. With \eqref{eq:c1c2_a} and \eqref{eq:c1c2_b}, we can determine the constants $C_1,C_2$ and up to second order in $\alpha$, we obtain the flux reflectivity  $\mathcal{R}_{\rm QED}$, for Quantum Electrodynamics (QED): 
\begin{eqnarray}
   \mathcal{R}_{\rm QED}^{\text{plasma}}  &&=  \Bigg{|}\frac{A_{\text{out}}}{A_{\text{in}}}\Bigg{|}^2~,\\
    &&=\frac{\alpha^2}{\mu^2} \Bigg|\frac{\Gamma\left(\frac{1}{2}-2 i \mu\right)}{\sqrt{2} \pi^{5/2} }
        +\frac{|m|^{2 i \mu}\Gamma(-2 i \mu) }{12 \pi  }\Bigg|^2~, \label{eq:reflectivity_plasma}\\
    &&\underset{\mu \gg 1}{\simeq} e^{-2 \pi \mu}\left(\frac{ \alpha^2 }{\pi^4 \mu^2}+ O \left[\frac{1}{\mu^{3}}\right]\right)\label{eq:reflectivity_plasma_large}~,\\
    &&= e^{-\frac{\omega}{T_H}}\left(\frac{ 4\alpha^2 T_H^2}{\pi^2 \omega^2}+ O \left[\frac{T_H^3}{\omega^{3}}\right]\right)\label{eq:reflectivity_plasma_large2}~,
\end{eqnarray}
where the flux reflectivity is evaluated for large values of $\omega/ T_H$ in \eqref{eq:reflectivity_plasma_large} and \eqref{eq:reflectivity_plasma_large2}. We see that we recover the Boltzmann factor of reflectivity for Quantum Black Holes that was also obtained and discussed extensively in earlier works \cite{Oshita:2019sat,Wang:2019rcf}. Furthermore,  we find an inverse square power-law dependence for large values of $\omega/ T_H$ which matches the calculation in Section \ref{sec:1-loop} up to second order correction in $\omega/ T_H$. Interestingly, the flux reflectivity obtained is independent on this class of interpolation in the large $\omega/ T_H$ limit. Taking the logarithm of reflectivity gives \begin{eqnarray}
   \ln \mathcal{R}_{\rm QED}^{\text{plasma}}  &&= \ln \left[\frac{\alpha^2}{\mu^2} \Bigg|\frac{\Gamma\left(\frac{1}{2}-2 i \mu\right)}{\sqrt{2} \pi^{5/2} }
        +\frac{|m|^{2 i \mu}\Gamma(-2 i \mu) }{12 \pi  }\Bigg|^2\right]~,\label{eq:logreflectivity_plasma}\\ 
   &&\underset{\omega/T_H \gg 1}{\simeq} -\frac{\omega}{T_H}-2 \ln \frac{\omega}{T_H}+\ln \frac{4 \alpha^2}{\pi^2}~,\label{eq:logreflectivity_plasma_large}
\end{eqnarray}

\begin{figure}[tbph]
	\centering
	\includegraphics[width=0.8\textwidth]{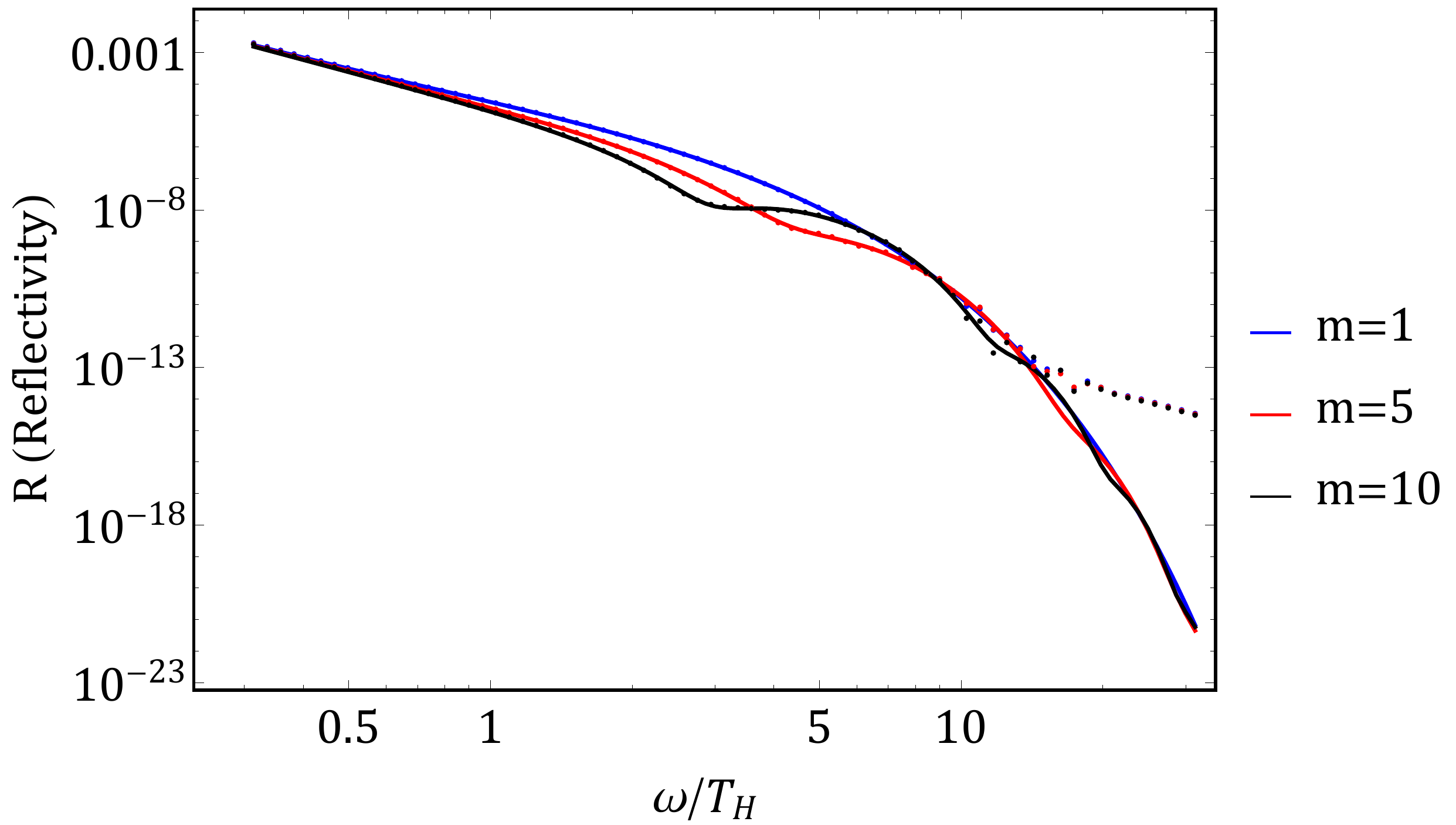}
	\caption{The reflectivity of incoming radiation by the relativistic Hawking plasma for various values of $m$. The dotted and solid lines are the numerical and analytical results respectively. The analytic expression is given in \eqref{eq:reflectivity_plasma}. There is a computational limit on the decimal points for large values of $\omega/ T_H$ and we have made a numerical plot up to that limit. We see that for small values of $\omega/ T_H$, the reflectivity obeys a power law. For large values of $\omega/T_H$, it is just the usual Boltzmann factor, corrected by a multiplicative $\omega^{-2}$ power-law dependence. The oscillatory term dies off as illustrated and shown in \eqref{eq:reflectivity_plasma_large2}. }
	\label{Fig:reflectivity-num}
\end{figure}

From \eqref{eq:reflectivity_plasma_large2}, we see that long wavelength photons can be scattered off the Hawking plasma due to significant interaction with the collective fermionic excitations. A photon is considered to have long wavelength if its wavelength is longer than that of the Hawking quanta at infinity. A long wavelength photon would thus overlap with several Hawking electrons/positrons upon reaching the horizon and hence, collective interactions must be taken into account. Short wavelength photons have their wavelength much shorter than the separation between the electrons upon reaching the horizon. Hence, no collective effects is present to modify earlier calculations done in \cite{Guo:2017jmi}. As a result of this, they are not reflected appreciably and fall into the horizon of the black hole. We numerically solve for \eqref{eq:wave_y} and the reflectivity is given by Figure \ref{Fig:reflectivity-num}.
\section{Reflectivity from 1-loop Correction to the Photon Propagator}
\label{sec:1-loop}
The dispersion relation \eqref{eq:dispersion_full} has a field theoretical description, where it can be described by the one-loop polarization tensor. The full derivation is given in \cite{Klimov:1982bv,Thoma:2008tb}. This motivates us to directly obtain the reflectivity by projection of the one-loop flat space photon propagator into Rindler modes. 

We are interested in obtaining the reflectivity for incoming radiation that experiences significant interaction with the Hawking plasma. The Hawking plasma is composed of thermal excitations of fermionic Rindler modes, and thus we need quantum field theory (QFT) in curved spacetime to describe its interactions. However, the Rindler space is just a wedge of the Minkowski spacetime seen by an accelerated observer and the two coordinates can be related directly, further simplifying the calculation. Doing the calculation in Minkowski coordinates will also avoid the ambiguity in the choice of vacuum, which is a notorious problem for QFT in curved spacetime.

\begin{figure}[tbph]
	\centering
	\includegraphics[width=0.4\textwidth]{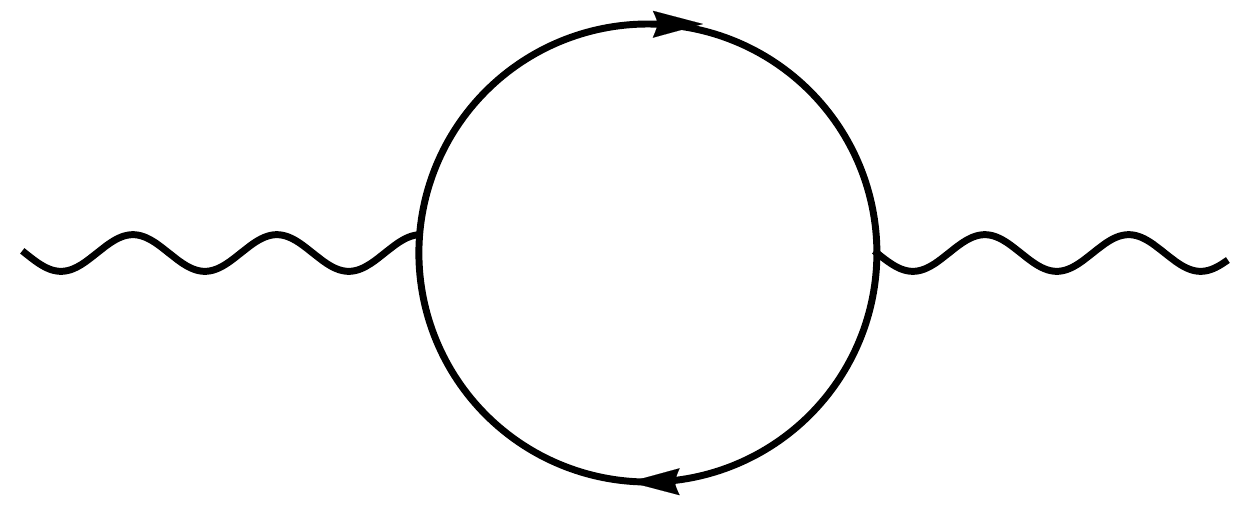}
	\caption{  The one-loop correction to the vacuum polarization diagram in QED. The fermionic loop allows the photon to interact with the virtual electron-positron pair during propagation. }
	\label{Fig:twopt_1loop}
\end{figure}

\begin{figure}[tbph]
	\centering
	\includegraphics[width=0.8\textwidth]{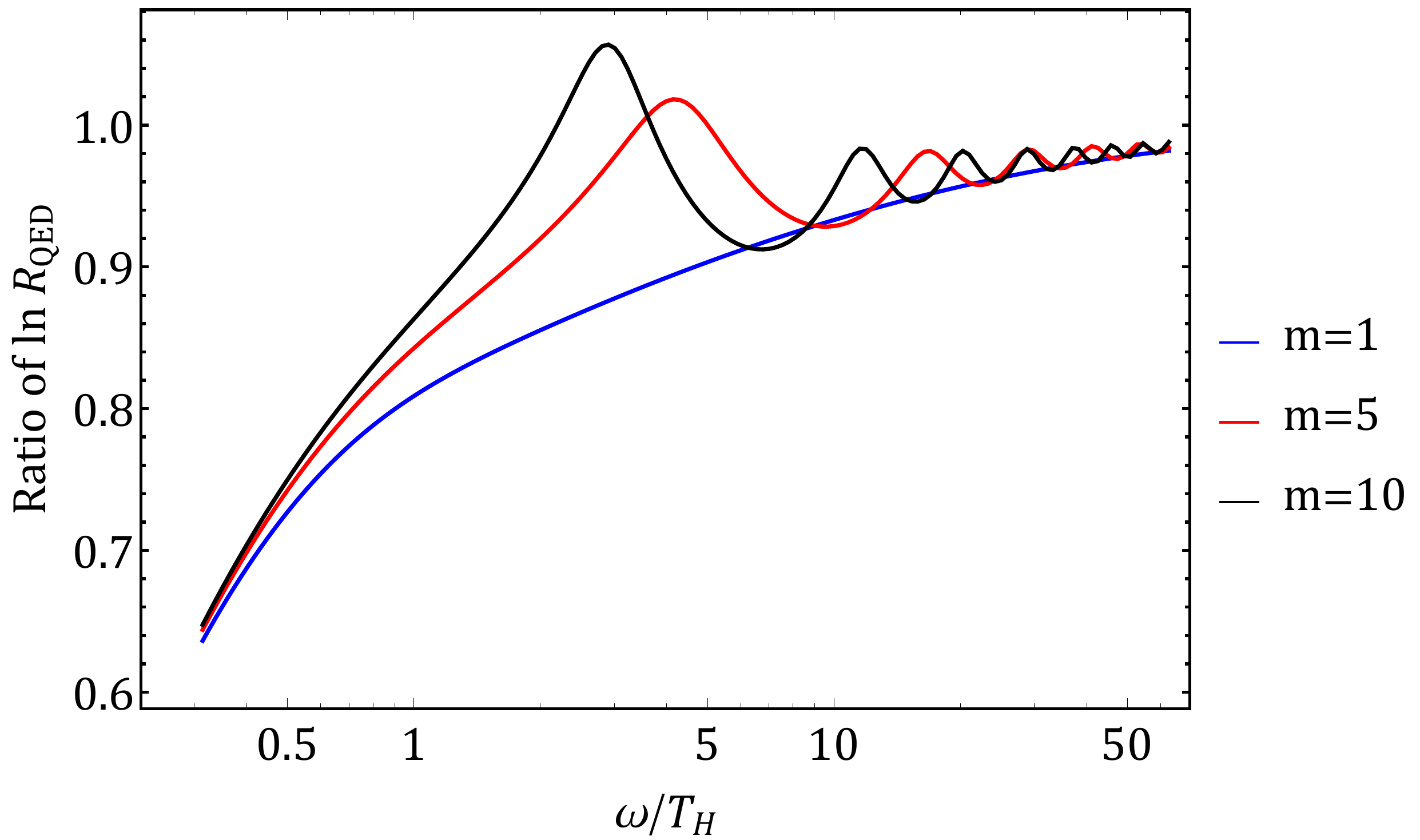}
	\caption{Ratio of log(1-loop reflectivity) to log(plasma reflectivity), from the two calculations presented here, using different interpolating functions (different m's). We can see that the oscillatory term dies off for large values of $\omega/T_H$ as demonstrated in \eqref{eq:logreflectivity_plasma_large}, and the two computations become asymptotically consistent.}
	\label{Fig:Rcompare}
\end{figure}

The relativistic quantum properties of charged fermions coupled to photons in a relativistic plasma is well described by the Dirac action:
\begin{equation}\label{eq:dirac}
    S_D = \int d^4 x \bar{\psi} \left( i\slashed{\partial}- e \slashed{A}-m\right)\psi~.
\end{equation}
The propagation of photons in the ambient $3+1$-dimensional space is described by the Maxwell action
\begin{equation}\label{eq:maxwell}
    S_M = -\frac{1}{4}\int d^4 x F_{\mu\nu}F^{\mu\nu}~.
\end{equation}
To the leading order in the fine structure constant $\alpha$, it is sufficient to consider the 1-loop correction of the photon propagator \cite{weinberg1995quantum}
\begin{equation}\label{eq:propagator}
    \Delta_{\mu\nu}^M(p) = \frac{\eta_{\mu\nu} + (\xi-1) \frac{p_\mu p_\nu}{p^2}}{(p^2+i \epsilon)(1-\pi^M(p^2))}~,
\end{equation}
where 
\begin{equation}
    \pi^M(p^2) = \frac{e^2}{2 \pi^2}\int_0^1 dx x(1-x) \ln \left(1+ \frac{p^2 x(1-x)}{m_e^2}\right)~.
\end{equation}
The one-loop correction to the photon propagator is illustrated in Figure \ref{Fig:twopt_1loop}. 
We consider the Feynman-'t Hooft gauge by taking $\xi= 1$ and the Fourier transform of the propagator is given by
\begin{equation}
    \Delta^M(X_M,X_M') = \int \frac{d^2 p}{(2 \pi)^2}e^{i p \cdot (X_M-X_M')}\Delta^M(p)~,\label{eq:photon_fourier}
\end{equation}
where $X_M = (T,X)$ is the 2D Minkowski coordinates. We only consider the Fourier transform in $(T,X)$ as we only consider the propagation of electromagnetic waves in the $(t,x)$ Rindler plane for obtaining the reflectivity \cite{Oshita:2019sat}. This is equivalent to assuming that the radial momentum is significantly blueshifted,  i.e., we neglect the transverse modes as a minor correction to the effective photon mass of the dispersion relation in \eqref{eq:effectmass} in the near-horizon limit. Assuming  the ingoing wave to be $A_R(\omega,-\mathbf{k})$ and outgoing wave to be $A_R(\omega,\mathbf{k})$ for a Rindler observer, a natural definition of flux reflectivity can be expressed in terms of the covariance of these quantum fields:
\begin{equation}\label{eq:reflectivity_def}
\mathcal{R}_{\rm QED}^{\text{1-loop}}  \equiv \Bigg{|}\frac{\langle A_R(\omega,\mathbf{k}) A^*_R(\omega,-\mathbf{k})\rangle}{\langle A_R(\omega,-\mathbf{k}) A^*_R(\omega,-\mathbf{k}) \rangle} \Bigg{|}^2,
\end{equation}
where
\begin{equation}
    \langle A_R(\omega,\mathbf{k})A^*_R(\omega',\mathbf{k'})\rangle = \int d^2 x e^{-i k \cdot x} \int d^2 x' e^{i k' \cdot x^{\prime}} \Delta^R(x,x') ~,
\end{equation}
is the Fourier transform in Rindler coordinates. We have further taken the photon propagator in Rindler $\Delta^R(x,x')$ to be equal to Minkowski $\Delta^M(X_M,X_M')$ in position space\footnote{This is justified as the two coordinate systems share the same $Y$ and $Z$ coordinates, and we are concerned with radially propagating photons.}.  For detailed computations regarding the flux reflectivity, please refer to Appendix \ref{appendix:1loop}. We only show the final result here, which is given by:
\begin{equation}
\begin{split}
    \mathcal{R}_{\rm QED}^{\text{1-loop}} &\simeq \Bigg{|} \frac{\langle A_R(\omega,|\omega|)A^*_R(\omega,-|\omega|)\rangle }{\langle A_R(\omega,-|\omega|)A^*_R(\omega,-|\omega|)\rangle} \Bigg{|}^2~,\\
   &\simeq \frac{\alpha^2 e^{-2\pi \mu}}{ 36 \pi^2 \mu^2}\label{eq:reflectivity_1loop}~,\\
   &= \frac{\alpha^2  T_H^2}{ 9 \omega^2}e^{-\frac{\omega}{T_H}}~, 
    \end{split}
\end{equation}
\begin{equation}
    \ln \mathcal{R}_{\rm QED}^{\text{1-loop}} = -\frac{\omega}{T_H}-2 \ln \frac{\omega}{T_H}+\ln \frac{ \alpha^2}{9}~,\label{eq:logreflectivity_1loop}
\end{equation}
where $ \alpha \equiv e^2/4 \pi$. We can see that the numerical coefficient of reflectivity in \eqref{eq:reflectivity_1loop} does not match with \eqref{eq:reflectivity_plasma_large}, while the power and exponent do. In principle, we can solve for the required interpolation in \eqref{eq:effectmass} to match the reflectivity in \eqref{eq:reflectivity_1loop}. However, it turns out that no exact interpolation can be obtained. Given that the numerical coefficients may be considered as a third order correction to the log(reflectivity) for large $\omega/ T_H$, we can interpret the mismatch as an artefact (or limitation) of the large $\omega/ T_H$ interpolation/approximation used in both derivations presented in Section \ref{sec:dispersion} and Appendix \ref{appendix:1loop}. A plot of ratio of log(reflectivity) obtained from two different methods, i.e. \eqref{eq:logreflectivity_plasma} and \eqref{eq:logreflectivity_1loop} is given by Figure \ref{Fig:Rcompare}. 
\section{Conclusion}
\label{sec:conclusions}
 Fuzzballs are the ``stringy'' version of Quantum Black Holes and are proposed to solve the well-known information paradox \cite{Mathur:2005zp}. Instead of the classical picture of an event horizon, we replace the black hole with a radiating surface composed of its microstate constituents. This solves the information paradox naturally as there is no horizon, and thus nothing can be trapped within the surface. 

In this work, we obtained the reflectivity of the incoming quanta from a fuzzball surface by considering the collective interaction with the relativistic $e^+e^-$ Hawking plasma in near-horizon Rindler geometry. We obtained the expression for reflectivity via two methods: using the photon dispersion relation within the thermal Hawking plasma, and the projection of the one-loop flat space photon propagator into Rindler modes. The two methods are indirectly related as there exists a field theoretical description for the plasma dispersion relation. Hence, it is intuitive that they show similar dependence in the end: The reflectivity is suppressed by the square of fine structure constant, a Boltzmann factor, and the inverse square of photon energy. We should note that the Boltzmann suppression of reflectivity for Quantum Black Holes was also obtained and discussed extensively in earlier works, e.g., \cite{Oshita:2019sat,Wang:2019rcf}, on more general grounds.

Our findings verifies the ``modified firewall conjecture'' in the context of Fuzzball proposal. The photons with energies comparable to Hawking temperature (i.e. wavelength comparable to horizon radius) can have significant interaction with the fermionic excitations of the fuzzball. However, in the large frequency limit, the photons can fall through the horizon unimpeded. In order to show this, we adopted the Unruh state in near-horizon Rindler's frame to model the populated fermionic excitations of the near-horizon Hawking plasma.

We found that for small frequencies, the reflectivity is enhanced by the collective fermionic interactions with the incoming photon. For $\omega \sim T_H$, the reflectivity is simply suppressed by $\alpha^2$ (i.e., much larger than $\alpha^2 (T_H/m_e)^2$ reported by \cite{Guo:2017jmi}). However, our results cannot be trusted down to arbitrarily small frequencies, as we are considering the relativistic regime of the plasma dispersion relation, i.e., $\omega  \gg m_\gamma \sim \sqrt{\alpha} T_H$. Nevertheless, it is interesting to see the reflectivity being directly enhanced by an inverse square dependence for small frequency modes. 

Let us now explore the connection between the electromagnetic reflectivity (or albedo) obtained here, i.e. \eqref{eq:reflectivity_plasma} and \eqref{eq:reflectivity_1loop}, and the more generic derivations of Boltzmann reflectivity in e.g., \cite{Oshita:2019sat,Wang:2019rcf}. While all these results have the same Boltzmann factor suppression, the 1-loop/plasma results found here are suppressed by an additional factor of $\sim (\alpha T_H/\omega)^2$. However, note that in this work, so far, we have ignored gravitational interaction, which is the primary ingredient in formation of a black hole. Similar to Fermi's four-point interaction, gravitational interaction is non-renormalizable at low energies, and thus its fine-structure constant scales as $\alpha_{\rm G} \sim \hat{E}_{\rm cm}^2/M^2_p$, where $\hat{E}_{\rm cm}$ is the centre of mass energy and $M_p$ is the Planck mass. For infalling photons of energy $\hat{E}_{\rm infalling}$, interacting with relativistic thermal electrons/positrons of temperature $T$, this yields:
\begin{equation}
    \alpha_{\rm G} \sim \frac{\hat{E}_{\rm infalling} T}{M^2_p} \Rightarrow \left(\frac{\alpha_g T_H}{\omega}\right)^2 \sim \frac{T^4}{M^4_p} 
    \to {\cal O}(1)~{\rm within~}l_p{\rm ~of~the~horizon},
\end{equation}
where we used $\omega/T_H = \hat{E}_{\rm infalling}/T $, for the blueshifted infalling photons. Unlike the electromagnetic reflection off the $e^+e^-$ plasma, which happens where $T \sim m_e$, the gravitational reflection is not significant until $T \sim M_p$ (i.e. a Planck proper length from the horizon). Moreover, note that at this point, the prefactor of the Boltzmann reflectivity (for gravitational reflectivity) becomes ${\cal O}(1)$ with no additional frequency dependence:
\begin{equation}
    {\cal R}_{\rm G} = {\cal O}(1) \times \exp\left(-\frac{\omega}{T_H}\right).
\end{equation}
Therefore, it is plausible to conclude that the simple Boltzmann reflectivity of \cite{Oshita:2019sat,Wang:2019rcf} may be the result of strong gravitational interactions with the Hawking plasma, while the perturbative electromagnetic interactions lead to $\alpha^2$ corrections computed here,  i.e. \eqref{eq:reflectivity_plasma} and \eqref{eq:reflectivity_1loop}. While the latter is subdominant, it is more robust and independent of near-horizon quantum gravity effects, such as vecro's and fuzzball bubbles \cite{Mathur:2020ely}. 

Let us conclude by a brief discussion of observational prospects. Now, given that we predict a significant albedo (or at least ${\cal R}_{\rm QED} \sim {\cal O(\alpha)} \sim 10^{-2}$) for low frequency photons, should we expect to see reflections from black hole horizons?  Unfortunately, the plasma frequency of the ambient interstellar medium provides a frequency cutoff of $f_p({\rm kHz}) \sim 10 \sqrt{n_e({\rm cm}^{-3})}$, which would be the primary hindrance for detecting low frequency radio waves; The Hawking frequency of a 10 solar mass non-spinning black hole is $10^2$ Hz, and only decreases for larger mass or spin. Nevertheless, the potential for observing similar quantum effects for radio pulsars orbiting black holes was entertained in \cite{2014MNRAS.445.3370P}, and deserves further exploration. Beyond radio astronomy, our findings provide further moral support for searches for gravitational wave echoes from quantum black holes, which are not hindered by interstellar plasma, and can be successfully carried out at frequencies comparable to those of Hawking radiation. This is an already  vibrant field of study (e.g., \cite{2016PhRvL.116q1101C,2017NatAs...1..586C,Abedi:2020ujo,PhysRevD.103.044028}). 

\acknowledgments

We would like to thank  Ruth Gregory, Samir Mathur, Paolo Pani, and Dan Wohns for comments on (different versions of) this manuscript, over the course of this project. This work was partially supported by Perimeter Institute, as part of the Perimeter Scholar International program. Research at Perimeter Institute is supported in part by the Government of Canada through the Department of Innovation, Science and Economic Development Canada and by the Province of Ontario through the Ministry of Colleges and Universities.


\bibliographystyle{JHEP}
\bibliography{Fuzzball}
\appendix
\section{Collisionless Boltzmann Equation in Rindler}
\label{appendix:CBE_Rindler}
In general relativity, collisionless matter moves along the geodesics, and Liouville's theorem implies the conservation of $f$ along geodesics \cite{rasio1989solving}
\begin{equation}
    \frac{D f_a(\mathbf{x_a}(\tau), \mathbf{p_a}(\tau))}{d \tau} = 0~,
\end{equation}
where the Liouville operator $D/d \tau$ is differentiation with respect to proper time along a geodesic:
\begin{equation}\label{eq:Liouville}
    \frac{D f_a(\mathbf{x_a}(\tau), \mathbf{p_a}(\tau))}{d \tau} \equiv \frac{d x_a^\alpha}{d \tau}\frac{\partial f_a}{\partial x_a^\alpha} + \frac{d p_a^\alpha}{d \tau}\frac{\partial f_a}{\partial p_a^\alpha} = 0~.
\end{equation}
We have the geodesics for a particle moving in the presence of electromagnetic field
\begin{equation}
    \frac{d^2 x_a^\alpha}{d \tau^2} + \Gamma^\alpha_{\mu \nu}\frac{d x_a^\mu}{d \tau}\frac{d x_a^\nu}{d \tau} = \frac{q_a}{m}F^{\alpha \nu}\frac{d x_a^\mu}{d \tau}g_{\mu\nu}~, 
\end{equation}
where $a=+/-$ as an indication for the charged particles. The Maxwell equations in curved spacetime read
\begin{eqnarray}
&& -\nabla_\alpha F^{\alpha\beta} = J^\beta = \sum_a g q_a n_a \frac{dx^\beta_a}{d\tau}~, \\
&& \nabla_\alpha F_{\beta\gamma} + \nabla_\beta F_{\gamma\alpha} + \nabla_\gamma F_{\alpha\beta} = 0~.
\end{eqnarray}
In order to solve the Collisionless Boltzmann Equation and Maxwell equation simultaneously, we can assume the distribution function $f$ to take the following form
\begin{equation}
    \begin{split}
        f_a(\mathbf{x}_a(\tau),\mathbf{p}_a(\tau)) &= f^{(0)}_a(\mathbf{x}_a(\tau),\mathbf{p}_a(\tau)) + f^{(1)}_a(\mathbf{p}_a(\tau))e^{i(-\omega t + \mathbf{k}\cdot \mathbf{r})}~, \nonumber\\
    &\simeq \frac{1}{e^{\frac{\sqrt{p^2 + m^2 e^{2 \kappa x}}}{T}}+1}+ f^{(1)}_a(\mathbf{p_a})e^{i(-\omega t + \mathbf{k}\cdot \mathbf{r})}~,
    \end{split}
\end{equation}
up to first order approximation. However, we weren't able to obtain plane-wave solutions for the Collisionless Boltzmann equation in Rindler \eqref{eq:Liouville}, and hence, we choose to obtain the effective photon mass by considering an effective interpolation \eqref{eq:effectmass}.
\section{One-Loop Photon Propagator: From Minkowski to Rindler}
\label{appendix:1loop}
We have the following relation between Rindler and Minkowski coordinates
\begin{equation}\label{eq:mink_rindler}
    \kappa(X+T) = e^{\kappa (x+t)} ~, \quad \kappa(X-T) = e^{\kappa (x-t)} ~.
\end{equation}
Directly computing the Jacobian $\Big|\frac{\partial(t,x)}{\partial(T,X)}\Big| = \frac{1}{\kappa^2(X^2-T^2)}$, and adopting \eqref{eq:mink_rindler}, we obtain
\begin{equation}\label{eq:photon_rindler}
\begin{split}
\langle A_R(\omega,\mathbf{k})A^*_R(\omega',\mathbf{k'})\rangle &= \int d^2 x e^{-i k \cdot x} \int d^2 x' e^{i k' \cdot x^{\prime }} \Delta^R(x,x') ~,\nonumber\\
&= \int_0^\infty dX\int_{-X}^X dT  \frac{1}{\kappa^2(X^2-T^2)}\left(\kappa(X+T)\right)^{-i \frac{k-\omega}{2 \kappa}}\left(\kappa(X-T)\right)^{-i \frac{k+\omega}{2 \kappa}} \nonumber\\
& \int_0^\infty dX' \int_{-X'}^{X'} dT'  \frac{1}{\kappa^2(X^{\prime 2}-T^{\prime 2})}\left(\kappa(X'+T')\right)^{i \frac{k'-\omega'}{2 \kappa}}\left(\kappa(X'-T')\right)^{i \frac{k'+\omega'}{2 \kappa}} \Delta^M(X_M,X_M')~,
\end{split}
\end{equation}
where $\Delta^M(X_M,X_M')$ is given in \eqref{eq:photon_fourier}. The Fourier transform is chosen such that there is no reflection of waves in the absence of interaction with the Hawking plasma, i.e.,
\begin{equation}
    \begin{split}
        \langle A_R(\omega,|\omega|)A^*_R(\omega,-|\omega|)\rangle^{(0)} = 0~.\label{eq:numzero}
    \end{split}
\end{equation}
at the zero order. Introducing the following coordinate transformation,
\begin{equation}\label{eq:UV}
U = \kappa(X-T)~,\quad V = \kappa(X+T)~,\quad p_u = p_1-p_0~, \quad
p_v = p_1+p_0~,
\end{equation}
and taking $\omega' = \omega, |k'| = -|k| = -|\omega| $, we subsequently have
\begin{equation}\label{eq:numzero2}
    \begin{split}
    &\langle A_R(\omega,|\omega|)A^*_R(\omega,-|\omega|)\rangle^{(0)} \\
        &=\frac{1}{2(4 \kappa^4)}\int \frac{d^2 p}{(2 \pi)^2}\int_0^\infty dU \frac{1}{U}U^{-i \mu}e^{i \frac{ p_v U}{2 \kappa}}\int_0^\infty dV \frac{1}{V}e^{i \frac{p_u}{2 \kappa}V} \\
        &\int_0^\infty dU'\frac{1}{U'}e^{-i \frac{p_u}{2 \kappa}U'}\int_0^\infty  dV'\frac{1}{V'}V^{\prime -i \mu}e^{-i \frac{p_v}{2 \kappa}V'}\frac{1}{p_u p_v+i \epsilon}~.
    \end{split}
\end{equation}
To evaluate \eqref{eq:numzero2}, we need to employ the following identities:
\begin{equation}
    \begin{split}
        \int_0^\infty dx\frac{1}{x}x^{-i \mu}e^{-i \frac{p x}{2 \kappa}} &= \left(\frac{i p}{2 \kappa}\right)^{i \mu}\Gamma(-i \mu)~, \\
        \int_0^\infty dx\frac{1}{x}e^{-i\frac{p x}{2 \kappa}} &= -\gamma -\ln\left(\frac{i p}{2\kappa}\right)~,
    \end{split}
\end{equation}
where $\gamma$ is the Euler's constant. We further compute
\begin{equation}
    \begin{split}
        \langle &A_R(\omega,|\omega|)A^*_R(\omega,-|\omega|)\rangle^{(0)} \\
        &= \frac{1}{2(4 \kappa^4)}\int \frac{d^2 p}{(2 \pi)^2}\left(\frac{p_v}{2 \kappa}\right)^{2 i \mu}\frac{\Gamma^2(-i \mu) }{p_u p_v+i \epsilon} \left[\gamma + \ln \left(\frac{i p_u}{2 \kappa}\right)\right]\left[\gamma + \ln \left(-\frac{i p_u}{2 \kappa}\right)\right]~, \\
        &= -\lim_{\Lambda  \rightarrow \infty}\frac{\Gamma^2(-i \mu)}{2(4 \kappa^4)}\int_0^\pi d\theta_1\int_0^\pi d\theta_2 \frac{1}{(2 \pi)^2}\left(\frac{\Lambda  e^{i \theta_2}}{2 \kappa}\right)^{2 i \mu} \left[\gamma + \ln \left(\frac{i \Lambda  e^{i \theta_1}}{2 \kappa}\right)\right]\left[\gamma + \ln \left(-\frac{i \Lambda e^{i \theta_1}}{2 \kappa}\right)\right]~, \\
        &= 0~,
    \end{split}
\end{equation}
where we have performed the contour integral over $p_u$ and $p_v$. $\Lambda $ is the UV cutoff in momentum. The flux reflectivity at first order correction in $\alpha$ is given by
\begin{equation}\label{eq:fluxR}
    \begin{split}
        \mathcal{R}_{\rm QED}^{\text{1-loop}} &= \Bigg{|}\frac{ \langle A_R(\omega,|\omega|)A^*_R(\omega,-|\omega|)\rangle^{(1)}}{\langle A_R(\omega,-|\omega|)A^*_R(\omega,-|\omega|)\rangle^{(0)} }\Bigg{|}^2~, \\
        &= \Bigg{|}\frac{\int d^2 p \int_0^\infty dU \frac{1}{U}U^{-i \mu}e^{i \frac{ p_v U}{2 \kappa}}\int_0^\infty dV \frac{1}{V}e^{i \frac{p_u}{2 \kappa}V} \int_0^\infty dU'\frac{1}{U'}e^{-i \frac{p_u}{2 \kappa}U'}\int_0^\infty  dV'\frac{1}{V'}V^{\prime -i \mu}e^{-i \frac{p_v}{2 \kappa}V'} \frac{\pi(p^2)}{p_u p_v + i \epsilon}}{\int d^2 p \int_0^\infty dU \frac{1}{U}e^{i \frac{ p_u U}{2 \kappa}}\int_0^\infty dV \frac{1}{V}V^{i \mu}e^{i \frac{p_v}{2 \kappa}V}\int_0^\infty dU'\frac{1}{U'}e^{-i \frac{p_u}{2 \kappa}U'}\int_0^\infty  dV'\frac{1}{V'}V^{\prime -i \mu}e^{-i \frac{p_v}{2 \kappa}V'} \frac{1}{p_u p_v + i \epsilon}}\Bigg{|}^2~. 
        \end{split}
\end{equation}
We first compute the denominator of \eqref{eq:fluxR}
\begin{equation}\label{eq:den}
    \begin{split}
        &\int d^2 p \int_0^\infty dU \frac{1}{U}e^{i \frac{ p_u U}{2 \kappa}}\int_0^\infty dV \frac{1}{V}V^{i \mu}e^{i \frac{p_v}{2 \kappa}V}\int_0^\infty dU'\frac{1}{U'}e^{-i \frac{p_u}{2 \kappa}U'}\int_0^\infty  dV'\frac{1}{V'}V^{\prime -i \mu}e^{-i \frac{p_v}{2 \kappa}V'} \frac{1}{p_u p_v + i \epsilon} \\
        &= \int d^2 p \frac{e^{\pi \mu} |\Gamma(i \mu)|^2}{p_u p_v+i \epsilon}\left[\gamma + \ln \left(\frac{i p_u}{2 \kappa}\right)\right]\left[\gamma + \ln \left(-\frac{i p_u}{2 \kappa}\right)\right]~, \\
        &= \int_{-\infty}^\infty dp_u \frac{(-i \pi)e^{\pi \mu} |\Gamma(i \mu)|^2}{p_u +i \epsilon}\left[\gamma + \ln \left(\frac{i p_u}{2 \kappa}\right)\right]\left[\gamma + \ln \left(-\frac{i p_u}{2 \kappa}\right)\right]~, \\
        &= -\pi e^{\pi \mu} |\Gamma(i \mu)|^2\lim_{\Lambda  \rightarrow \infty}\int_0^\pi d\theta \left[\gamma + \ln \left(\frac{i \Lambda  e^{i \theta}}{2 \kappa}\right)\right]\left[\gamma + \ln \left(-\frac{i \Lambda  e^{i \theta}}{2 \kappa}\right)\right]~, \\
        &\underset{\Lambda  \rightarrow \infty}{\simeq} -\pi^2 e^{\pi \mu}|\Gamma(i \mu)|^2 \ln^2 \Lambda ~.
    \end{split}
\end{equation}
We further perform the computation for the numerator
\begin{equation}\label{eq:numone}
    \begin{split}
        &\int d^2 p \int_0^\infty dU \frac{1}{U}U^{-i \mu}e^{i \frac{ p_v U}{2 \kappa}}\int_0^\infty dV \frac{1}{V}e^{i \frac{p_u}{2 \kappa}V} \int_0^\infty dU'\frac{1}{U'}e^{-i \frac{p_u}{2 \kappa}U'}\int_0^\infty  dV'\frac{1}{V'}V^{\prime -i \mu}e^{-i \frac{p_v}{2 \kappa}V'} \frac{\pi(p^2)}{p_u p_v + i \epsilon} \\
        &= \int d^2p \left(\frac{p_v}{2 \kappa}\right)^{2 i \mu}\frac{\Gamma^2(-i \mu) }{p_u p_v+i \epsilon} \left[\gamma + \ln \left(\frac{i p_u}{2 \kappa}\right)\right]\left[\gamma + \ln \left(-\frac{i p_u}{2 \kappa}\right)\right]\pi(p^2)~, \\
        &\underset{p^2 \gg m_e^2}{\simeq} \int d^2p \int_0^1 dx\left(\frac{p_v}{2 \kappa}\right)^{2 i \mu}\frac{\Gamma^2(-i \mu) }{p_u p_v+i \epsilon} \left[\gamma + \ln \left(\frac{i p_u}{2 \kappa}\right)\right]\left[\gamma + \ln \left(-\frac{i p_u}{2 \kappa}\right)\right]\frac{e^2}{2 \pi^2}x(1-x) \ln \left(\frac{p_u p_v x(1-x)}{m_e^2}\right)~, \\
        &= \lim_{\Lambda  \rightarrow \infty} -\int_0^1 dx \int_0^\pi d\theta_1 \int_{0}^\pi d\theta_2\left(\frac{\Lambda  e^{i \theta_2}}{2 \kappa}\right)^{2 i \mu}\Gamma^2(-i \mu) \left[\gamma + \ln \left(\frac{i \Lambda  e^{i \theta_1}}{2 \kappa}\right)\right]\left[\gamma + \ln \left(-\frac{i \Lambda  e^{i \theta_1}}{2 \kappa}\right)\right] \\
        &\times \frac{e^2}{2 \pi^2}x(1-x) \ln \left(\frac{ x(1-x)}{m_e^2}\Lambda ^2 e^{i \theta_1}e^{i \theta_2}\right)~, \\
        &\underset{\Lambda \rightarrow \infty}{\simeq} -\frac{i e^2}{24 \mu}\left(\frac{\Lambda }{2 \kappa}\right)^{2 i \mu}\Gamma^2(-i \mu)\ln^2 \Lambda~.
    \end{split}
\end{equation}
Combining the calculation in \eqref{eq:den} and \eqref{eq:numone}, we obtain the expression for the flux reflectivity
\begin{equation}
    \begin{split}
        \mathcal{R}_{\rm QED}^{\text{1-loop}}&\simeq \frac{e^4 e^{-2 \pi \mu}}{576\pi^4 \mu^2}~, \\
        &= \frac{\alpha^2 e^{-2 \pi \mu}}{36\pi^2 \mu^2}~, \\
        &= \frac{\alpha^2  T_H^2}{ 9 \omega^2}e^{-\frac{\omega}{T_H}}~,
    \end{split}
\end{equation}
where we have defined the fine structure constant $\alpha \equiv e^2/4 \pi$ in natural units. The cutoff in momentum that we imposed drops out naturally in the calculation for flux reflectivity. 
\end{document}